\documentclass{sig-alternate-2013}

\usepackage[utf8]{inputenc}
\usepackage{enumitem}
\usepackage[hyphens]{url}
\usepackage[pdftex,urlcolor=black,colorlinks=true,linkcolor=black,citecolor=black]{hyperref}

\usepackage{CJKutf8}

\usepackage[usenames,dvipsnames,svgnames,table]{xcolor}
\usepackage{fancyvrb}
\usepackage{relsize}
\usepackage{listings}
\usepackage{verbatim}
\newcommand{\defaultlistingsize}{\fontsize{8pt}{9.5pt}}
\newcommand{\inlinelistingsize}{\fontsize{8pt}{11pt}}

\newcommand{\listingsize}{\defaultlistingsize}
\RecustomVerbatimCommand{\Verb}{Verb}{fontsize=\inlinelistingsize}
\RecustomVerbatimEnvironment{Verbatim}{Verbatim}{fontsize=\defaultlistingsize}
\let\oldurl\url
\renewcommand{\url}[1]{\inlinelistingsize\oldurl{#1}}

\lstdefinelanguage{JavaScript}{
  keywords={console, log, addEventListener, onmessage, alert, push, typeof, new, true, false, catch, function, return, null, catch, switch, var, if, in, while, do, else, case, break},
  keywordstyle=\bfseries,
  ndkeywords={class, export, boolean, throw, implements, import, this},
  ndkeywordstyle=\color{darkgray}\bfseries,
  identifierstyle=\color{Maroon},
  sensitive=false,
  comment=[l]{//},
  morecomment=[s]{/*}{*/},
  commentstyle=\color{ForestGreen},
  stringstyle=\color{Blue},
  morestring=[b]',
  morestring=[b]"
}

\usepackage{color}
\definecolor{grey}{RGB}{130,130,130}
\newcommand{\linewrap}{\raisebox{-.6ex}{\textcolor{grey}{$\hookleftarrow$}}}

\usepackage{color}

\permission{Copyright is held by the International World Wide Web Conference Committee (IW3C2). IW3C2 reserves the right to provide a hyperlink to the author's site if the Material is used in electronic media.}
\conferenceinfo{WWW'14 Companion,}{April 7--11, 2014, Seoul, Korea.} 
\copyrightetc{ACM \the\acmcopyr}
\crdata{978-1-4503-2745-9/14/04. \\
http://dx.doi.org/10.1145/2567948.2576948}

\begin{document}
%

\title{Bots vs. Wikipedians, Anons vs. Logged-Ins}

\numberofauthors{1}

\author{
\alignauthor
Thomas Steiner\titlenote{Thomas Steiner's second affiliation is \emph{Université de Lyon, CNRS Université Lyon~1, LIRIS, UMR5205, F-69622}}\\
       \affaddr{Google Germany GmbH}\\
       \affaddr{ABC-Str.~19, 20354 Hamburg, Germany}\\
       \email{tomac@google.com}
}

\maketitle
\begin{abstract}
Wikipedia is a~global crowdsourced encyclopedia
that at time of writing is available in 287 languages.
Wikidata is a~likewise global crowdsourced knowledge base
that provides shared facts to be used by Wikipedias.
In the context of this research, we have developed
an application and an underlying
Application Programming Interface~(API) capable of monitoring
realtime edit activity of all language versions
of Wikipedia and Wikidata.
This application allows us to easily analyze edits
in order to answer questions such as
``Bots \emph{vs.}\ Wikipedians, who edits more?'',
``Which is the most anonymously edited Wikipedia?'',
or ``Who are the bots and what do they edit?''.
To the best of our knowledge,
this is the first time such an analysis
could be done in realtime for Wikidata \emph{and}
for really \emph{all} Wikipedias---large and small.
Our application is available publicly online at the URL
\url{http://wikipedia-edits.herokuapp.com/},
its code has been open-sourced under the Apache~2.0 license.
\end{abstract}




\section{Introduction}

The free online encyclopedia Wikipedia%
\footnote{Wikipedia: \url{http://www.wikipedia.org/}}
was formally lau\-nched on January 15, 2001
by Jimmy Wales and Larry Sanger,
albeit the fundamental wiki technology and the underlying concepts are older.
Wikipedia's initial role was to serve
as a~collaborative platform for draft articles for Nupedia,
an earlier, also free online encyclopedia that was exclusively edited by experts.
What happened in practice was that Wikipedia rapidly overtook Nupedia
as there was no peer-review burden
and it is now a~globally successful Web encyclopedia
available in 287 languages%
\footnote{List of Wikipedias by size:
\url{http://meta.wikimedia.org/wiki/List_of_Wikipedias}}
with overall more than 30 million articles.%
\footnote{Wikipedia statistics: \url{http://stats.wikimedia.org/}}

\paragraph{The First Wikipedia Bots}

Wikipedia bots are computer programs
with the purpose of automatically editing Wikipedia.
After occasional smaller-scale tests,
the first large-scale bot operation
was started in October 2002 by Derek Ramsey,%
\footnote{History of Wikipedia bots:
\url{http://bit.ly/History_Bots}}
who created a~bot to add a~large number
of articles about United States towns.
The generated articles used a~uniform
text template, so that all articles
followed the same writing style.
Today, bots are not only used to generate articles,
but also to fight vandalism and spam,
and many more automatable tasks.%
\footnote{Wikipedia bots by purpose: \url{http://en.wikipedia.org/wiki/Category:Wikipedia_bots_by_purpose}}

\paragraph{The Knowledge Base Wikidata}

As Wikipedia is a~truly global effort,
sharing non-language-dependent facts
like population figures centrally
in a~know\-ledge base makes a~lot of sense
to facilitate international article expansion.
Wikidata\footnote{Wikidata: \url{http://www.wikidata.org/}}~\cite{vrandecic2012wikidata}
is a~free knowledge base that can be read
and edited by both humans and bots.
The knowledge base centralizes access to
and management of structured data,
such as references between Wikipedias
and statistical information that can be used in articles.
Controversial facts such as borders in conflict regions
can be added with multiple values and sources,
so that Wikipedia articles can,
dependent on their standpoint, choose preferred values.

\paragraph{Contributions}

The contributions of this paper are twofold.
First, we have developed an application
and released its source code as open-source
that allows for realtime monitoring of all 287~Wikipedias and Wikidata.
Second, we have permanently made available a~publicly useable
Application Programming Interface~(API) that our application
is based upon and that we invite other interested parties to use.

\section{Methodology and Tools}

\paragraph{Wikipedia Recent Changes}
\label{sec:wikipedia-recent-changes}

Whenever a~human or bot changes an article
of any of the 287 Wikipedias,
a~change event gets communicated by a~chat bot
over the Wikimedia IRC server (\url{irc.wikimedia.org}),%
\footnote{Raw IRC feeds of recent changes:
\url{http://meta.wikimedia.org/wiki/IRC/Channels\#Raw_feeds}}
so that parties interested in the data
can listen to the changes as they happen%
~\cite{steiner2013mjnomore}.
For each language version, there is
a~specific chat room following the pattern
\texttt{"\#" + language + ".wikipedia"}.
An exception from this pattern is the room
\texttt{\#wikidata.wikipedia} for the language-independent
knowledge base Wikidata~\cite{vrandecic2012wikidata}.
A~sample original chat message with the components separated
by the asterisk character \texttt{`*'}
announcing a~change to an article
can be seen in the following.
\texttt{"[[Keep Calm and Carry On]] http://en.wikipedia.org/w/index.php?diff=585806152\\\&oldid=585805943 * 74.197.171.148 * (+14) /* Paro\-dies */"}.
The components are \emph{(i)}~article name, \emph{(ii)}~revision URL,
\emph{(iii)}~editor handle,
\emph{(iv)}~change size and description.

\paragraph{Server-Sent Events}

Server-Sent Events~\cite{hickson2012sse}
defines an API for using an HTTP connection
to receive push notifications from a~server
in the form of DOM events.
Therefore, on the server side, a~script generates messages
of the MIME type \texttt{text/event-stream}
in an event stream format that can be seen
in \autoref{code:sse-server}.
The required event payload is in the \texttt{data:} field,
events can optionally be typed via a~proceeding \texttt{event:} field.
Consecutive events are separated by two line breaks.
The \texttt{EventSource} interface enables Web applications
to listen to pushed events from a~server over the HTTP protocol.
On the client side, using this API consists of creating
an \texttt{EventSource} object and registering event listeners,
as can be seen in \autoref{code:sse-client}.

\begin{figure}[h!]
  \begin{lstlisting}[caption={Server-Sent Event of type ``enedit''
    (formatted for legibility, \texttt{data:} allows no line breaks)},
    label=code:sse-server, language=JavaScript]
event: enedit
data: {
  "article": "Golden_Globe_Award_for_Best_§\linewrap§
      Actress_-_Motion_Picture_Musical_or_Comedy",
  "editor": "en:86.150.237.133",
  "isBot": false,
  "language": "en",
  "diffUrl": "http://en.wikipedia.org/w/api.php?§\linewrap§
      action=compare&torev=585820379&fromrev=§\linewrap§
      585776128&format=json"
}
  \end{lstlisting}

  \vspace{-1.5em}
  \begin{lstlisting}[caption={\texttt{EventSource}
    object with event listener},
    label=code:sse-client, language=JavaScript]
  // connect to SSE stream and attach listener
  var source = new EventSource('/sse');
  source.addEventListener('enedit', function(e) {
    console.log((JSON.parse(e.data)).article);
  }, false);
  \end{lstlisting}

  \vspace{-0.5em}
  \fbox{\includegraphics[width=\linewidth]{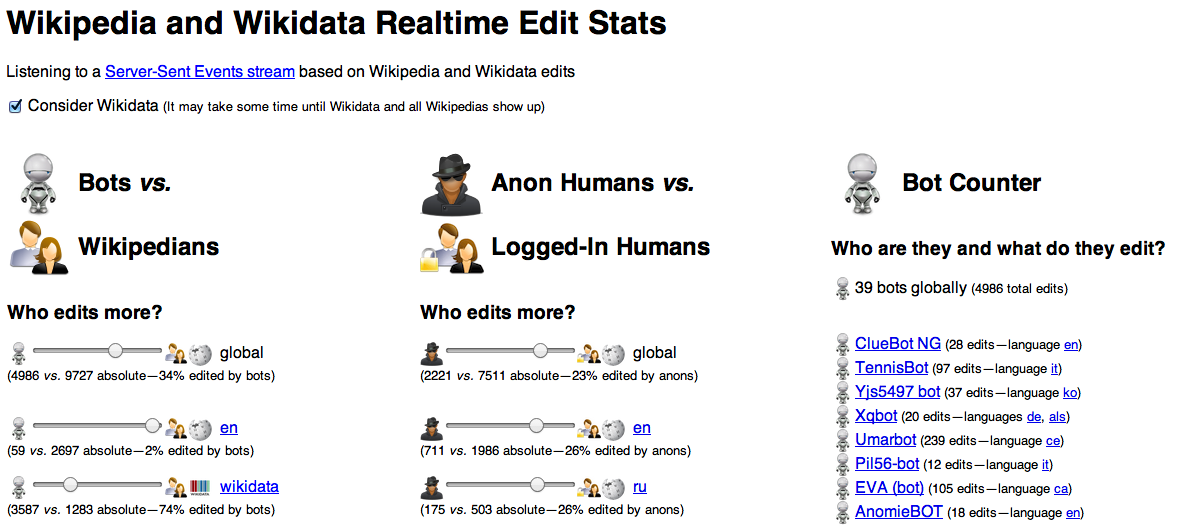}}
  \caption{Screenshot of the application (cropped)}
  \label{fig:screenshot}
\end{figure}

\paragraph{Implementation Details}

Our application is based on a~Server-Sent Events API
that we have implemented in Node.js,
a~server side JavaScript software system
designed for writing scalable Internet applications.
Using Martyn Smith's Node.js IRC library,%
\footnote{Node IRC:
\url{https://github.com/martynsmith/node-irc}}
we listen for Wikipedia and Wikidata edit events
and send Server-Sent Events whenever we detect one.
Our API is available publicly online
at the URL \url{http://wikipedia-edits.herokuapp.com/sse}
and open for third parties to use.
On the client side, we have registered generic event handlers for events
pushed by the API and keep track of edit statistics over time.
A~screenshot can be seen in \autoref{fig:screenshot}.

\section{Preliminary Results}

In a~first iteration, we have observed all 287~Wikipedias and Wikidata
during the observation period November~4
to~6, 2013.
Already during this short period,
overall exactly 3,805,185~edit events occurred.
Our application updates in realtime,
which allows us to detect when relative figures,
\emph{i.e.}, percentages of bots \emph{vs.}\ Wikipedians
and anonymous \emph{vs.}\ logged-in humans start to converge.
This was the case after about half of the observation period.
At the end of the observation,
from all 287~Wikipedias and Wikidata,
exactly 260~($\sim90.3\%$) were edited.
Our toolset being publicly available,
interested parties can run longer analyses at will.

\section{Conclusions}

We have introduced an open-source application
and underlying API for the realtime monitoring
of all 287~Wiki\-pedias including Wikidata
and have analyzed more than 3.8~million edit events
to get a~better understanding
of the relations of logged-in \emph{vs.}\ anonymous edits
and edits made by bots \emph{vs.}\ edits made by humans.
Concluding, we have contributed a~useful toolset
and performed a~preliminary global study
with interesting insights that
is the first in a~series
of future studies and applications by us and others.

\bibliographystyle{abbrv}
\bibliography{wbsc21p-steiner}
\balancecolumns
\end{document}